\documentclass[twocolumn,showpacs,preprintnumbers,superscriptaddress,amsmath,amssymb,prl]{revtex4}

\usepackage{graphicx,epsfig,color} 
\usepackage{dcolumn}
\usepackage{bm}
\usepackage{amsmath}
\usepackage{indentfirst}
\usepackage{url}

\newcommand{\be}{\begin{equation}}
\newcommand{\ee}{\end{equation}}
\newcommand{\ba}{\begin{eqnarray}}
\newcommand{\ea}{\end{eqnarray}}

\begin{document}

\preprint{APS preprint}

\title{Empirical Tests of Zipf's law Mechanism In Open Source Linux Distribution}
 
\author{T. Maillart}
\affiliation{Chair of Entrepreneurial Risks, Department of
Management, Technology and Economics, ETH Zurich, CH-8001 Zurich,
Switzerland \\.}

\author{D. Sornette}
\affiliation{Chair of Entrepreneurial Risks, Department of
Management, Technology and Economics, ETH Zurich, CH-8001 Zurich,
Switzerland \\.}

\author{S. Spaeth}
\affiliation{Chair of Strategic Management and Innovation,  Department of
Management, Technology and Economics, ETH Zurich, CH-8001 Zurich, Switzerland}

\author{G. Von Krogh}
\affiliation{Chair of Strategic Management and Innovation,  Department of
Management, Technology and Economics, ETH Zurich, CH-8001 Zurich, Switzerland}

\date{\today}

\begin{abstract}

The evolution of open source software projects in Linux distributions offers a remarkable example of a growing complex self-organizing adaptive system, exhibiting Zipf's law over four full decades. We present three tests of the usually assumed ingredients of stochastic growth models that have been previously conjectured to be at the origin of Zipf's law: (i) the growth observed between successive releases of  the number of in-directed links of packages obeys Gibrat's law of proportional growth; (ii)  the average growth increment of the number of in-directed links of packages over a time interval $\Delta t$  is proportional to  $\Delta t$, while its standard deviation is proportional to $\sqrt{\Delta t}$; (iii) the distribution of the number of in-directed links of new packages appearing in evolving versions of Debian Linux distributions has
a tail thinner than Zipf's law, with an exponent which converges to the Zipf's law value $1$ as the time $\Delta t$ between releases increases.

\end{abstract}



\maketitle

Complex adaptive systems in nature and society often exhibit scale-free properties,
either in their self-organizing fractal geometry and/or their self-similar statistical distributions.
Among the many such characteristics, Zipf's law plays a particular role as one of the few quantitative reproducible regularities found in the social sciences.
Zipf's law usually refers to probability density functions $p(x)$ of some stochastic variable $x$, usually a size or frequency, exhibiting the power law dependence 
\begin{equation}
p(x) \sim 1 / x^{1+\mu}~~~{\rm with}~~\mu = 1~. 
\label{wbtwr}
\end{equation}
Initially formulated as a rank-frequency relationship quantifying the relative commonness of words in natural languages \cite{Zipf}, Zipf's law accounts remarkably well for the distribution of city sizes \cite{Gabaix99} as well as firm sizes \cite{Simon-Bonini,IS77,Axtell} all over the world. Recently, Zipf's law has also been found in Web access statistics and Internet traffic characteristics  \cite{Huberman1,AlbertBara02} as well as in bibliometrics, informetrics, scientometrics, and library science (see \cite{Huberman2} and references therein).  

Starting with Yule \cite{Yule} and Schumpeter \cite{Schumpeter1934},  it is now recognized that  there are important links between such size distributions and growth.  On this basis, Simon \cite{Simon1} articulated a simple mechanism for Zipf's law based on 
Gibrat's law of proportionate effect \cite{Gibrat1931} implemented in a stochastic growth model
with new entrants. A modern formulation of Gibrat's law is that growth is a random
process, with successive stochastic realizations of the growth rates that are independent of the size of the entity (city, firm, website popularity and so on). In the context of the distribution of firm sizes, Simon \cite{Simon1}  modified Gibrat's model by accounting for the entry of new firms over time as the overall industry grows. 
This model has recently been rediscovered under the name ``preferential attachment'' to explain the scale-free networks found in social communities, the world-wide web, or networks of proteins reacting with each other in biological cells \cite{BaraAlbert99,AlbertBara02}. 
But the existence of new entrants in the growth process is just one of the many different 
additional ingredients complementing Gibrat's law that yields Zipf's law \cite{Champernowne,Kesten73,Sornette98,Gabaix99}.  

While several works have tested Gibrat's law directly (and its deviations) in various contexts, 
and have conjectured on its relevance to explain Zipf's law, we present
the first fully consistent empirical study showing that the usually assumed ingredients of stochastic growth models are indeed present in a system exhibiting Zipf's law. For this, we provide an empirical analysis of the growth of an operating system (Debian Linux) based on open source softwares.
Large Linux distributions  typically contain tens of thousands of connected packages, including the operating system and applications, which form a complex web of inter-dependencies. A measure of the ``centrality'' of a given package is the number of other packages that call it in their routine, a measure we refer to as the number of in-directed links or connections that other packages have to a given package. We find that the distribution of in-directed links of packages in successive Debian Linux distributions precisely obeys Zipf's law  over four orders of magnitudes.  We then verify explicitly that the
growth observed between successive releases of  the number of in-directed links of packages obeys Gibrat's law with a good approximation. As an additional critical test of the stochastic growth process, we confirm empirically that the average growth increment of the number of in-directed links of packages over a time interval $\Delta t$  is proportional to  $\Delta t$, while its standard deviation is proportional to $\sqrt{\Delta t}$, as predicted from Gibrat's law implemented in a standard stochastic growth model. In addition, we verify that the distribution of the number of in-directed links of new packages appearing in evolving version of Debian Linux distributions has a tail thinner than Zipf's law, confirming that Zipf's law in this system is controlled by the growth process. 

The Linux Kernel was created in 1991 by Linus Torvalds as a clone of the proprietary Unix operating system \cite{kernel,torvalds1999le}, and was licensed under GNU General Public License. Its code and open source license had immediately a strong appeal to other developpers who contributed to its further development. Quickly, the community  of open source developers started to run other open source programs on this new operating system. In 1993, Debian Linux \cite{debian}  became the first non-commercial successful general distribution of an open source operating system. While continuously evolving, it remains up to the present the ``mother'' of a dominant Linux branch, competing with a growing number of derived distributions (Ubuntu, Dreamlinux, Damn Small Linux, Knoppix, Kanotix, and so on). 

\begin{figure}
\centerline{\epsfig{figure=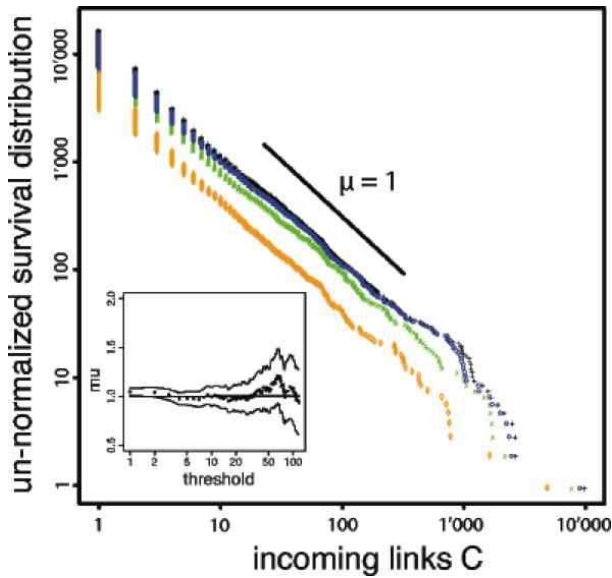,angle=0,width=8cm, scale=3}}
\caption{(Color Online) Log-log plot of the number of packages in four Debian Linux Distributions with more than $C$ in-directed links. The four Debian Linux Distributions are
Woody (19.07.2002) (orange diamonds), Sarge (06.06.2005) (green crosses), Etch (15.08.2007) (blue circles), Lenny (15.12.2007) (black $+$'s). The inset shows the Maximum Likelihood estimate (MLE) of the exponent $\mu$ together with two boundaries defining its $95\%$ confidence interval (approximately given by $1 \pm 2/\sqrt{n}$, where $n$ is the number of data points using in the MLE), as a function of the lower threshold. The MLE has been modified from the standard Hill estimator to take into account the discreteness of $C$.}
\label{fig:zipf}
\end{figure}

From a few tens to hundreds of packages (474 in 1996 (v1.1)), Debian has expanded to include more than about 18'000 packages in 2007, with many intricate dependencies between them, that can be represented by complex functional networks. Debian offers a remarkable example of a growing complex self-organizing adaptive system \cite{MyersSoftware}. Its evolution is recorded by a chronological series of stable and unstable releases: new packages enter, some disappear, others gain or lose connectivity. Here, we study the following sequence of Debian releases: Woody: 19.07.2002;  Sarge: 0.6.06.2005; Etch: 15.08.2007; Lenny (unstable version): 15.12.2007; several other Lenny versions from 18.03.2008 to 05.05.2008 in intervals of 7 days.

Figure \ref{fig:zipf} shows the number of packages in the first four successive versions of Debian Linux with more than $C$ in-directed links, which is nothing but the un-normalized complementary cumulative (or survival) distribution of package numbers of in-directed links. Zipf's law is confirmed over four full decades, for each of the four releases. Notwithstanding the large modifications between releases and the multiplication of the number of packages by a factor of three between Woody and Lenny, the distributions shown in Fig.\ref{fig:zipf} are all consistent with Zipf's law.  It is remarkable that no noticeable cut-off or change of regimes occurs neither at the left nor at the right end-parts of the distributions shown in Fig.\ref{fig:zipf}. Our results extend those conjectured in Ref.~\cite{Challet2004p3340} for Red Hat Linux. By using Debian Linux, which is better suited for the sampling of projects than the often used SourceForge collaboration platform, we avoid biases and gather unique information only available in an integrated environment \cite{Spaeth2007}. 

To understand the origin of this Zipf's law, we use the general framework of 
stochastic growth models, and we track the time evolution of a given package
via its number $C$ of in-directed links connecting it to other packages within Debian Linux.
The increment $dC$ of the number of in-directed links to a given package over a small
time interval $dt$ is assumed to be the sum of two contributions, defining
a generalized diffusion process:
\begin{equation}
dC= r(C)~dt + \sigma(C) ~dW~,
\label{eq:stoc_proc}
\end{equation}
with $r(C)$ is the average deterministic growth of the in-directed link number, $\sigma(C)$ is the standard 
deviation of the stochastic component of the growth process and $dW$
is the increment of the Wiener process (with $\langle dW \rangle =0$ and 
$\langle dW^2 \rangle =dt$ where the brackets denote performing the statistical average). 
Zipf's law has been shown to arise under a variety of conditions associated with
Gibrat's law. The simplest implementation of Gibrat's law writes that both 
$r(C)$ and $\sigma(C)$ are proportional to $C$, 
\begin{equation}
r(C)=r \times C~,~~~~~~\sigma(C)= \sigma \times C~, 
\label{tkhtbw}
\end{equation}
with proportionality coefficients $r$ and $\sigma$ obeying the following inequality 
$r < \sigma$. This later inequality expresses that the proportional growth 
is dominated by its stochastic component \cite{MalSaiSor07}. Accordingly, the heavy tail structure of Zipf's law can be thought of as the result of large stochastic multiplicative excursions.
The rest of the letter is devoted to testing and validating this model.

\begin{figure}
\centerline{\epsfig{figure=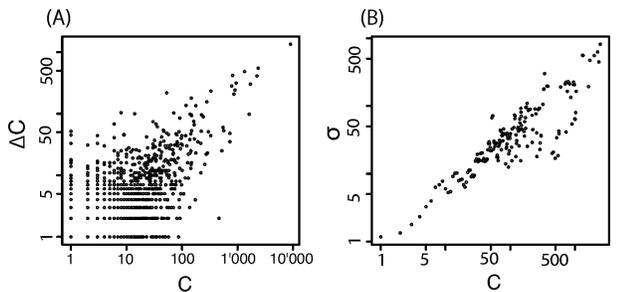,angle=0,width=8cm, scale=3}}
\caption{Left panel: Plots of $\Delta C$ versus $C$ from the Etch release (15.08.2007) to the latest Lenny version (05.05.2008) in double logarithmic scale. Only positive values are displayed. The linear regression $\Delta C= R \times C + C_0$  is significant at the $95\%$ confidence level, with a small value $C_0=0.3$ at the origin and $R=0.09$. Right panel: same as left panel for the standard deviation of $\Delta C$.
}
\label{fig:reglin}
\end{figure}

First, we measure the time evolution of the in-directed links of all packages in the successive
Debian releases,  by retrieving the network of dependencies following the methodology explained
in Ref. \cite{Spaeth2007}. For packages which are common to successive releases, we find that their connectivity, measured for instance by their number $C$ of in-directed links, increases on average albeit with considerable fluctuations. Consider for instance the update from Etch (15.08.2007) to the latest Lenny version (05.05.2008). For each package $i$ which is common to these two versions, we measure the increment $\Delta C_i$ of the number $C_i$ of in-directed links to that package from Etch to the latest Lenny version. The left panel of Fig.\ref{fig:reglin} plots these increments $\Delta C_i$ as a function of $C_i$. This figure is typical of the results obtained on the increments  $\Delta C_i$ between other pairs of Debian releases. The scatter plot confirms the existence of an approximate proportionality between $\Delta C_i$ and $C_i$, especially for the largest $C_i$ values, in agreement with the first equation of (\ref{tkhtbw}). The right panel of Fig.\ref{fig:reglin} shows the standard deviation of $\Delta C$ as a function of $C$, confirming the second equation of (\ref{tkhtbw}). These two panels are nothing but direct evidence of Gibrat's law for package connectivities, which constitutes an essential ingredient of stochastic growth models of Zipf's law  \cite{Champernowne,Kesten73,Sornette98,Gabaix99}. 
Notice that the large scatter decorating the approximate proportionality between $\Delta C_i$ and $C_i$ observed in Fig.~\ref{fig:reglin} and quantified in the right panel of Fig.\ref{fig:reglin}  is an essential ingredient for Zipf's law to appear \cite{MalSaiSor07}.

\begin{figure}
\centerline{\epsfig{figure=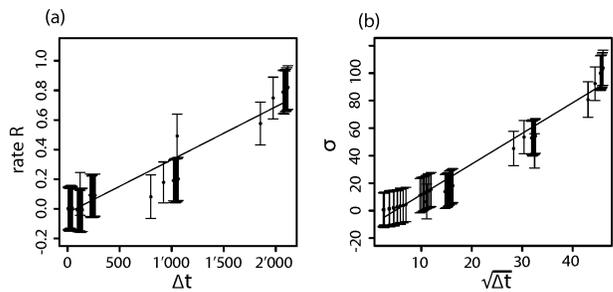,angle=0,width=8cm, scale=5}}
\caption{Dependence of $R(\Delta t)$ and $\Sigma(\Delta t)$ defined respectively by
$R(\Delta t) \equiv \langle \Delta C / C \rangle$ and
(\ref{hwtwqqergr}) as a function of their time interval $\Delta t$ for the 66 time intervals that can be
formed between all the Debian releases in our database (which includes the four major Debian releases from 19.07.2002 to 15.12.2007 as well as the several Lenny releases from 18.03.2008 to 05.05.2008 in intervals of 7 days). The error bars show the 95\% confidence intervals, obtained by shuffling 1000 times the linear regression residuals. The straight lines represent the best linear fits. The existence of a genuine linear dependence of $R$ as a function of $\Delta t$ cannot be rejected ($p<0.05$) and has a high signifiance level (square of correlation coefficient ${\cal R}^2=0.93$). The regression of $\Sigma$ versus $\sqrt{\Delta t}$ enjoys the same high statistical confidence ($p<0.05$ and ${\cal R}^2=0.97$).}
\label{fig:bsdt}
\end{figure}

We then combine (\ref{eq:stoc_proc})  and (\ref{tkhtbw}) to predict that, over a not too large time interval $\Delta t$, (i) the average growth rate $R(\Delta t) \equiv \langle \Delta C / C \rangle$ should be given by  
\begin{equation}
R(\Delta t) = r \times \Delta t~,
\label{firstpr}
\end{equation}
and (ii) the standard deviation of the growth rate 
\begin{equation}
\Sigma(\Delta t) \equiv \langle [\Delta C / C]^2 \rangle^{1 \over 2}
\label{hwtwqqergr}
\end{equation}
should be equal to 
\begin{equation}
\Sigma(\Delta t) = \sigma \times \sqrt{\Delta t}~.
\label{firstpr2}
\end{equation}
This last result derives from the properties of the Wiener process increments $dW$.
We test these two predictions (\ref{firstpr}) and (\ref{firstpr2}) as follows. Out of the four major Debian releases from 19.07.2002 to 15.12.2007 as well as the several Lenny releases from 18.03.2008 to 05.05.2008 in intervals of 7 days, 66 different time intervals can be formed. For each time interval, we
calculate the average growth rate defined by $R(\Delta t) \equiv \langle \Delta C / C \rangle$ and its standard deviation  defined by (\ref{hwtwqqergr}). Technically, we estimate
$R(\Delta t)$ (respectively $\Sigma(\Delta t)$) as the slope (respectively the
standard deviation of the residuals) of the linear regression of $\Delta C$ as a function of $C$.
This method allows us to construct confidence bounds by bootstrapping (we reshuffle 1000 times the linear regression residuals).  The left (resp. right) panel of figure \ref{fig:bsdt} shows the 66 values of $R(\Delta t)$ (resp. $\Sigma(\Delta t)$) as a function
of their corresponding time interval $\Delta t$ (resp. square-root of $\Delta t$), providing a strong validation of the stochastic growth model  (\ref{eq:stoc_proc})  and (\ref{tkhtbw}).

We now address the question of how the increase of the number of packages interacts with 
the growth process of the number of links between packages. This issue has been
considered in the context of firms (see Ref. \cite{MalSaiSor07} for a detailed presentation and
summary of the literature), and applies to the case of packages as follows.
Most stochastic growth models of firms based on Gibrat's principle attempt to derive the distribution of the cross-section of firm sizes directly from the distribution of the asset value of a single firm as a function of time. Indeed, many models start with the implicit or explicit assumption that the set of firms was born at the same origin of time. This approach is mathematically equivalent to considering that the universe is made only of one single entity. Therefore, the distribution of firm sizes can reach a steady-state if and only if the distribution of the asset value of a single firm reaches a steady state, which is  counterfactual. A more correct model is to take into account the fact that firms do not appear all at the same time but are born according to a more or less regular flow of newly created firms. Competing with the birth process, firms also disappear at a surprisingly high rate. Similarly, the evolution of successive Debian releases is punctuated by additions and deletions of many packages. 
For instance, at the release of the latest stable release (Lenny, 15.12.2007), $885$ packages disappeared, partly merged, or were renamed while $2983$ packages appeared compared to the precedent release. Clearly, the dynamics of the connectivity between packages depends on the birth as well as demise of packages. Therefore, the stochastic growth model (\ref{eq:stoc_proc})  
must be supplemented by a model of the birth and death of packages. For this, we use Saichev et al.  \cite{MalSaiSor07}'s approach, who showed that Gibrat's law of proportionate growth does not need to be strictly satisfied in the presence of  the birth and death of entities following the stochastic growth process (\ref{eq:stoc_proc}): as long as the volatility $\Sigma(\Delta t)$ defined by (\ref{hwtwqqergr})
increases asymptotically proportionally to $C$ and that the instantaneous growth rate increases not faster than the volatility, the distribution of sizes follows Zipf's law. This suggests that the occurrence of very large firms in the distribution of  sizes described by Zipf's law is more a consequence of random growth than systematic returns. Likewise, in particular for packages with large connectivities, volatility can dominate over the instantaneous growth rate. 

\begin{figure}
\centerline{\epsfig{figure=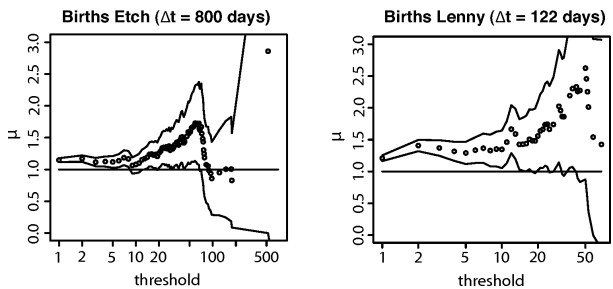,angle=0,width=8cm, scale=5}}
\caption{ The right panel shows that the exponent $\mu$ of the distribution of $C$'s
of new packages appearing between successive unstable Lenny releases separated by one week is a power law with exponent $\mu \simeq 1.5$; the left panel show that the same power law has a smaller exponent closer to $1$ as one considers the new packages appearing between two more distant releases. We have verified that this effect is systematic in our database.
The exponents $\mu$ are obtained by maximum likelihood, adapted to the discreteness of $C$ values.
The thin lines defined the $95\%$ confidence intervals.}
\label{fig:threh_evol}
\end{figure}

Figure \ref{fig:threh_evol} verifies that the distribution of the numbers $C$ of in-directed links of newly born packages has a tail thinner than Zipf's law,  and converges progressively to Zipf's law as the time elapsed between two releases increases, reflecting the increasing impact of the stochastic multiplicative growth process.  This confirms that Zipf's law results indeed from the stochastic multiplicative growth process at the level of individual packages in the presence of the birth-death of packages. 

\end{document}